\input epsf.sty
\documentstyle[11pt,paspconf]{article}

\begin{document}

\title{Simulating Cosmic Structure at High Resolution: Towards a
Billion Particles?}  

\author{H.M.P. Couchman}
\affil{Department of Physics and Astronomy, University of Western
Ontario, London, Ontario, N6A 3K7, Canada}

\begin{abstract}
Cosmic structure simulations have improved enormously over the past
decade, both in terms of the resolution which can be achieved, and
with the addition of hydrodynamic and other techniques to formerly
purely gravitational methods. This is an informal, and perhaps
idiosyncratic, overview of the state and progress of cosmological
simulations over this period.

I will discuss the strategies that are used to understand cosmic
structure formation and the requirements of a successful simulation. 
The computational demands of cosmological simulations in general will
be highlighted as will specific features of the various different
algorithms that are used. The incessant push for greater resolution
has lead to an increasing use of parallel computers. The
desirability---or otherwise---of this trend will be discussed
together with alternative techniques which sidestep the na\"\i ve drive for
greater resolution.
\end{abstract}


\section{Introduction}

One of the key issues of contemporary cosmology is to understand the
formation and evolution of cosmic structures; from the pattern of
clusters and galaxies on the largest scales down to the formation of
galaxies themselves. In the standard model cosmic structure grew
from a spectrum of low amplitude fluctuations present at the epoch of
recombination. The form of this spectrum is at present unknown, although
a number of experiments are placing constraints on the fluctuation
amplitude at recombination on large scales. Early next century
satellite missions, such as COBRAS/SAMBA and MAP, promise to tightly
constrain the recombination-epoch spectrum as well as a number of
other significant cosmological parameters. It is the spectrum of
fluctuations at recombination which forms the initial conditions for
studies of structure formation.

Whilst the fluctuation amplitude is small post-recombination growth
is linear; spectral modes grow independently and proportionately to
the universal expansion (assuming that the Universe is close to flat
at the relevant epoch). Presently-observed structures range from the
quasi-linear such as the large scale network of filaments and voids
and superclusters to the highly non-linear such as galaxies and galaxy
clusters.  Whilst linear growth of small fluctuations is well
understood and analytically tractable, only approximate methods and
perturbation theory or simple ``toy'' models are available for
studying non-linear growth. Making the connection between the initial
fluctuation spectrum and presently observed structures, a task central
to our understanding of the post-recombination universe, requires
numerical simulation. The need for numerical simulation becomes even
more pressing when we add the hydrodynamic component necessary for an
understanding of gas clusters and, especially, dissipation in galaxies.

The paper is laid out as follows. I begin with a pictorial description
of the growth of the fluctuation spectrum and the transition from
linear to non-linear growth for a generic class of spectra in which
bound objects progress from small to large. I then discuss the
requirements of a numerical simulation in terms of the way in which we model
the universe and the force and mass resolution necessary for various
aspects of structure formation. Following a description of the
motivation for the use of particle methods, the various different
algorithms that have been developed are summarized. A brief historical
overview of the achievements of simulation methods is then
presented. The push towards the use of supercomputers for numerical
simulation is considered and I will present a case study of this
endeavour. I conclude with a view of the problems posed by very large
simulations and speculate about the future.

\section{Growth of Fluctuations}

Many popular contemporary cosmological models consider spectra in
which the mass variance is a decreasing function of scale. Such
spectra can explain a number of observational features, such as, for
example, the fact that clusters have formed more recently than
galaxies.

%[164 172 487 587]
\begin{figure}
\centerline{\epsfxsize=3.7truein \epsfbox{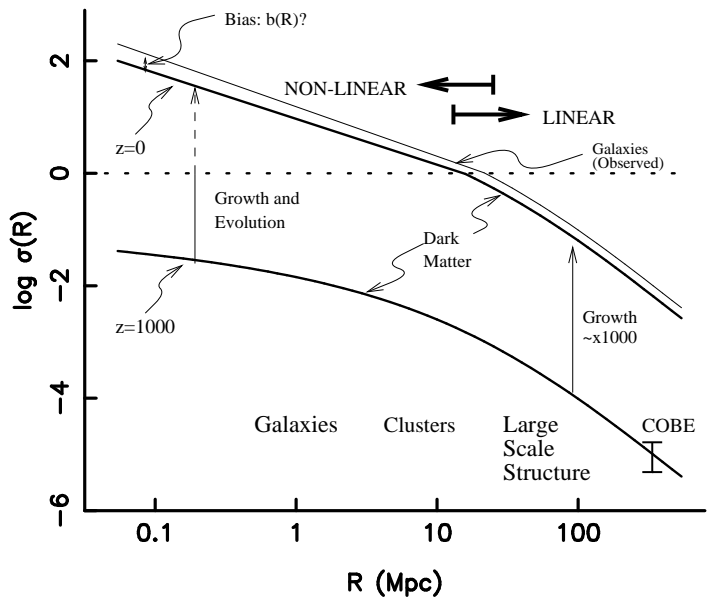}}
\caption{A schematic view of the post-recombination growth of the
fluctuation spectrum, here represented as the mass variance as a
function of scale. Whilst the mass variance is less than unity, growth
is linear, for larger amplitudes reliable prediction requires
numerical simulation.}
\end{figure}

Figure~1 shows a pictorial view of the growth of the mass variance as
a function of scale. (The precise form of the spectrum is unimportant
only the decrease to large scales is significant for this discussion.)
The figure illustrates in a generic way at what scales and epochs we
encounter non-linear gravitational evolution. Detailed understanding
of structures on scales for which $\sigma(R)\ga 1$ at any epoch
requires numerical simulation.

Key questions which will benefit from a numerical approach include:

\begin{itemize}
\item The large-scale distribution of galaxies and their ``bias'' relative
to the underlying gravitationally dominant dark matter component.
\item The formation and evolution of clusters; the behaviour of the
intra-cluster medium and the distribution and evolution of the galaxy
population.
\item Understanding the structure of the universe out to the redshifts
of quasars; Lyman--$\alpha$ forest etc.
\item The formation and influence of the first bound objects at
redshifts between 10 and 40.
\item Galaxy formation itself---the ``Holy Grail'' of
post-recombination cosmology.
\end{itemize}

The value of numerical simulation is that it allows us to surmount the
obstacle of non-linear gravitational and hydrodynamic evolution.  Not
only will we be able to investigate the formation and evolution of
presently observed structure and probe its immediate precursors,
but we will be able to make the crucial connection with the initial
fluctuation spectrum.

\section{Simulation Requirements}

Cosmological simulations have considered a number of different aspects
of structure formation. These have ranged from studies of the large
scale distribution of matter, without specific reference to galaxies,
to studies of individual clusters and galaxies. I will focus on the
so-called ``Grand Challenge'' problem of simulating the formation and
distribution of galaxies in the cosmological context. This study is
well motivated: galaxies form the fundamental building blocks of the
universe. With the advent of huge galaxy redshift surveys such as the Sloan
Digital Sky Survey and the Two Degree Field we will have an enormous
observational database with which to interpret and constrain
theoretical models. The relationship of the galaxies that we observe
to the overall matter distribution in the universe is, however,
uncertain. It is crucial that we understand the nature of any bias
that may be present. It is only through detailed numerical simulation
of the formation and evolution of galaxies (which are highly
dissipated objects) that we can hope to achieve a detailed
understanding of cosmic structure.

Broadly speaking the requirements of such a simulation are that we
model a representative piece of the universe over a sufficient range
of scales with accurate forces and time integration schemes. These
are, of course, very general requirements which apply to any well
conceived numerical simulation. One of these desirable qualities,
accurate time integration, I will mention only in passing. Typically
second order leapfrog or low order Runge--Kutta or
Predictor--Corrector schemes are employed. This is largely a
consequence of the need to minimize the storage of extra quantities
for a huge number of particles whose orbits must be integrated,
coupled with the expense of the force calculation. That it is
acceptable to use such low order schemes is a result of studying
gravitational collapse. Because of the instability of the system we
cannot hope to follow individual particle orbits accurately but are
only concerned that we adequately model the properties of bound
objects in a statistical sense. Note that this is in stark contrast to
the extreme care which is exercised in the integration of orbits in
the Solar System as described elsewhere in this volume.

The issue of the boundary conditions to apply to the simulation will
of course depend upon the precise nature of the problem being
considered. I will return briefly to this issue towards the end of the
paper. It suffices at present to note that, for the study of the
large-scale distribution of galaxies, it is convenient to chose a
cubic simulation volume with triply periodic boundary conditions which
is large enough, subject to the limitations of available resolution,
to represent a fair sample of the universe.

For the remainder of this section I will concentrate on two aspects of
resolution; force and mass resolution. Gaining higher and higher
resolution has been a primary focus of much of the effort expended in
cosmological simulations over the last decade. I will attempt to
justify this trend and at the same time give the motivation for the
use of particle methods.

\subsection{Force Resolution}

The ``top hat'' model is a useful simplification of the collapse of an
initially overdense region. An overdense spherically symmetric region
within an otherwise homogeneous universe will have an expansion and
contraction corresponding to the evolution of a super-critical
universe. In practice, inhomogeneities and external tides will
generate non-radial motions which will prevent the region collapsing
to a singularity and cause it to virialize at roughly half its
turnaround radius. At virialization, in a flat universe, the
overdensity of the object will be nearly 200, and this will increase
as the background continues to expand. An object formed at a redshift
of 4 would have an overdensity of 10$^4$ at redshift zero, even
without dissipation.

If we are to properly model cosmic structure it is essential to follow
and retain bound objects at increasing density contrast. An instance
of the importance of this would be in an investigation of
substructure; galaxies within a cluster for example. Since a
cosmological simulation typically models a comoving region of the
universe these very large density contrasts are present in the
simulation. Popular choices for modelling the wide range of densities
and varied geometries that occur are Lagrangian particle methods.
Although some early simulations represented each galaxy with a single
particle it is now much more common to model the matter density as a
collisionless fluid which is approximated by an ensemble of softened
particles. An important question is then the choice of softening. To
strictly maintain the fluid limit the softening should never be less
than the mean interparticle separation, and yet this precludes
accurate tracking of highly over-dense regions within a comoving
volume. It is more realistic to demand that the softening be of order
the mean interparticle separation {\it within a bound object}, and
thus maintain the collisionless fluid approximation locally. Since
objects collapsing at different epochs have different physical
densities a spatially variable softening is indicated. Most workers,
however, chose a fixed spatial softening for the gravitational
interaction, although as the dynamic range of simulations increases
this question will need to be revisited.

The art of choosing the appropriate softening lies in negotiating between
the unwanted effects of two-body relaxation if the value is too small,
and over-merger of ``fluffy'' substructure if the value is too
large. Typically, realistic simulations of galaxies in a cosmological
context, for example, require force-softenings at least an order of
magnitude smaller than the mean interparticle separation.

\subsection{Mass Resolution}

As noted above, a huge amount of effort has been expended in running
simulations with ever larger particle number. I give a simple
order-of-magnitude estimate for a suitable particle number for the
large-scale galaxy distribution Grand Challenge problem.

To model a fair sample of the universe requires that we simulate a
cube of at least 100\thinspace Mpc on a side.  A volume of this size
will contain roughly 10$^4$ bright galaxies. The fraction of the
overall matter density residing in galaxies is perhaps one
percent. Allowing for 100 particles per galaxy halo suggests that a
simulation should contain at least 10$^8$ resolution elements or
particles. Only now is it becoming possible to run simulations with
this particle number and typically the simulations which have been run
have not had high resolution forces. The above argument, whilst
simplistic, demonstrates clearly the need for very high resolution.

A further leap in the required resolution becomes necessary when one
wishes to model the dissipative component of galaxies. Suppose that
one uses Smoothed Particle Hydrodynamics (SPH; Gingold and
Monaghan~1977) to model the gaseous content of galactic haloes. SPH,
which fits well with the particle codes, requires at least 10$^3$
particles per galaxy to reliably follow the cooling and collapse of
gas in dark haloes, and likely considerably more. This implies a
simulation with $\sim10^9$ particles.

Not all investigations require modelling of huge volumes of the
universe. Two examples will illustrate the requirements in other
situations. 

\begin{itemize}
\item {\bf Galaxy Formation} $4\times10^6$ particles in a 10\thinspace
Mpc region $\rightarrow$ 1\thinspace kpc resolution.
\item {\bf Clusters} $4\times10^7$ particles in a 50\thinspace
Mpc region $\rightarrow$ 3\thinspace kpc resolution.
\end{itemize}

Simulations with such huge numbers particles are not without drawbacks
of course. The storage for positions and velocities alone for 10$^8$
particles is 2.4\thinspace Gb for each time-slice (32bit words), and
roughly 50\% greater for the storage of all quantities in an SPH
simulation.  Nonetheless, the problem is well motivated and, at least
for the problem under consideration, these sorts of particle numbers
are being considered and the first such runs performed. Various
alternatives to the straightforward approach of greater particle
number are mentioned in the final section.

\section{N-Body Algorithms}

The task which must be accomplished then, is the computation of the
gravitational interaction of a very large number of particles. Whilst
conceptually simple this problem has consumed a vast amount of effort
and programming skill and has led to a number of different
approaches. The primary popular approaches are described in outline
below. All of these, apart from the first, calculate the gravitational
potential by dividing the field into two components; the near field
which has high frequency content and the far field which is
approximated to some order by a given basis expansion. Assuming that
the force at an arbitrary point is required to a specified accuracy,
it can be argued heuristically that these methods all scale as $O(N\ln
N)$, where $N$ is the particle number.

\begin{itemize}
\item {\bf Particle--Particle (PP)} Forces on each particle are
accumulated by directly summing over all neighbours. Whilst this method
is very robust it is $O(N^2)$ and is hence limited to fewer than
$\sim10^{3\hbox{--}4}$ particles. (It is worth noting that this method
has received a new lease of life in cosmology as a result of the
development of application-specific chips which compute the Newtonian
force in hardware.)
\item {\bf Tree methods} These methods are the logical successors to
the PP method although historically they appeared after the standard
grid-based techniques described next. These methods take account of
the fact that to a given accuracy a group of particles distant from
the force calculation point may be approximated by a low order
multipole expansion thus avoiding the expensive sum over all particles
in the group. Various types of tree have been used to organize the
data in such a way that the decision as to whether or not a cell of
particles need be subdivided for a particular force calculation may be
made efficiently. For details see Barnes \& Hut~(1986), Hernquist \&
Katz~(1989). A number of variants exist. The Fast Multipole Method of
Greengard \& Rokhlin~(1987) is similar in spirit to this approach. It
will not be described further here as its use in cosmology has been
very limited.
\item{\bf Grid-Based Methods} These methods sample the density field
with a (usually) uniform grid. Poisson's equation is then solved on
this grid using one of a number of the fast solvers that are
available, usually either an FFT or multigrid. The potential obtained
corresponds to the far field component as the grid cannot represent
frequencies higher than the Nyquist limit. The Particle-Mesh (PM)
force is typically augmented by a short range contribution summed over
near neighbours, thus allowing high resolution spatial forces. The
method can become grossly inefficient as clustering develops, however,
and the short-range direct sum becomes dominant. This problem has been
alleviated in the Adaptive P$^3$M (Couchman~1991; AP$^3$M) code by
using a hierarchy of adaptive meshes in regions of high particle
density.
\end{itemize}

Most workers now use either a version of the Tree code or a grid-based
method (primarily P$^3$M, or variant, using Fast Fourier Transforms),
although a number of other techniques have been, or are being
developed. These include Grid-Tree hybrids in which a tree replaces
the expensive direct sum in P$^3$M and other grid techniques in which
a true grid refinement strategy is used.

There are a number of pros. and cons. for each of the primary
methods. Tree codes are robust, the cycle time does not vary much
between light and heavy clustering and they provide a flexible data
structure which makes implementation of individual particle timesteps,
parallelization and inclusion of SPH relatively straightforward. Their
primary drawbacks are that they use a substantial amount of memory,
$\sim25\hbox{--}35$ words per particle, and, for cosmology, the
necessary implementation of periodic boundary conditions via the
Ewald~(1921) summation method can be cumbersome.

P$^3$M has the advantage that it is very fast under conditions of
light clustering, has automatically periodic boundary conditions when
using an FFT to solve for the grid potential and uses relatively
little memory; $\sim8\hbox{--}10$ words per particle. A severe
disadvantage is that clustering can dramatically slow the algorithm
unless an efficient scheme for relieving the PP work in highly
clustered regions is implemented. Although, the use of regular uniform
grids throughout the calculation leads to efficiency in terms of
memory use and execution speed under light particle clustering, it
results in a less general data structure than a tree, for example,
which makes inclusion of individual particle timesteps, as well as
some aspects of parallelization, more difficult.

\section{Hydrodynamics}

The addition of a hydrodynamic component to gravitational cosmological
codes permits the investigation of many new aspects of structure
formation. Of particular importance has been the study of the hot
intra-cluster medium.  The number density and distribution of clusters
provide important constraints on the primordial spectrum. More
recently several ambitious attempts have being made upon the prize of
galaxy formation itself, in which correct modelling of dissipative
processes is crucial.

Perhaps the most straightforward technique for adding hydrodynamics to
a particle code is Smoothed Particle Hydrodynamics (SPH). In this
method thermodynamic quantities are carried by particles and the value
of the fluid is approximated at any point by interpolating values from
nearby particles. Because of the ease of integrating this method with
gravitational particle methods it was the first widely used
hydrodynamic method in cosmology. Implementations of SPH in both Tree
codes (Hernquist \& Katz~1989) and P$^3$M (Evrard~1990) exist.

More recently a number of alternative schemes have been imported into
cosmological codes. In particular Eulerian codes have become
popular. These offer a number of potential advantages over SPH, in
particular efficient shock capturing schemes greatly improve on the
number of resolution elements necessary to model a shock; in 1-D two
or three cells for an Eulerian code and 6 particles for SPH. These
methods are also well suited to the addition of magneto-hydrodynamics
and radiation fields. In order for these codes to follow the large
density contrasts that arise, adaptive mesh refinement techniques are
essential, and these are now beginning to be used. Other schemes
include semi-Lagrangian methods and very recently an unstructured
finite element method has been described (Xu~1996). In all of these
codes particles carry the gravitational mass.

\section{Achievements}

In this section I will summarise the improvements that have occurred
in the field of cosmological simulations over the last decade, very
briefly list the areas in which they have played a significant
role in advancing the subject, and outline the present status of the
field.

Although it is clear that cosmological simulations are an essential
tool for cosmologists studying structure formation, they have not
proved to be the central mechanism for generating new ideas or insight
about the post-recombination universe. The physical ideas are simple
and well understood; provided that we can quantify the collective
behaviour of particles, simulations are helpful in understanding the
behaviour and interaction of these physical processes in highly
complicated situations. In this respect cosmological simulations have
a different focus and utility than simulations of say, galactic
dynamics. In the latter (well studied) case the galaxy may be
idealised to a degree which allows detailed investigation of, for
example, modes of a galactic disc. In cosmological simulations the
focus is rather different. First we are typically trying to model a
realization of an initial random field (albeit with known initial
spectrum) which lacks the symmetries of the galaxy study. Second,
analytic theory beyond first order, so far, has quite limited
predictive power. A reasonable expectation is that simulation and
theory will play a mutually supportive role in improving our knowledge
in this area.

A convenient milepost to the development of what might be termed
``modern'' cosmological simulation methods is the pioneering work of
Davis, Efstathiou, Frenk and White in the mid 1980s (see Efstathiou et
al.~1995 for a description of numerical methods). This work, following
from earlier work by Efstathiou and Eastwood drew the P$^3$M code of
Hockney and Eastwood~(1988) from the plasma physics arena into the
cosmological community and represented what might be termed the first
high resolution cosmological simulations with $\sim3\times10^4$
particles. 

With the availability of efficient algorithms, increases in computer
speed and memory size led rapidly to higher resolution simulations. By
the late 80s $\sim5\times10^5$ particle P$^3$M simulation were run and
this number increased to $\sim2\times10^6$ by 1990. At present it is
possible to run routinely $\sim2\times10^7$ particle AP$^3$M
simulations. Differing memory and cpu requirements mean that at a
given time somewhat larger PM, or smaller Tree-code, simulations could
be performed. Hydrodynamic simulations both because of their larger
memory requirement and the necessity of using a smaller timestep are
generally at present limited to a few million particles.

A common criticism of cosmological simulation efforts has been that
they are merely a method of generating attractive pictures. It is
certainly true that, so far, there have been only a few useful
statistical measures used to express the collective behaviour of the
large number of particles in a cosmological simulation. It is
worth bearing in mind the comments made above, however, concerning the
likely role that simulations will play in improved understanding of
structure formation. The following items of progress are essentially
all cases in which simulations have checked analytic models or
predictions. This is only a sample of the areas in which simulations
have played a key role but it is indicative of the way in which
simulations have been, and are likely to continue to be, employed.

\begin{itemize}
\item Checks of perturbation theory.
\item Testing the validity of the Press--Schechter~(1974) model, which
attempts to model the could-in-cloud problem.
\item Seeking a universal scaling function to describe the non-linear
mass auto-correlation as a function of the initial linear auto-correlation
(Hamilton et al.~1991).
\item Determining cluster distributions and abundances
\item Attempts to understand galaxy bias through an investigation of
the distribution of putative galaxy haloes in non-hydrodynamic simulations.
\item Understanding the intra-cluster gas.
\item Modelling Lyman-$\alpha$ clouds and Lyman-$\alpha$ forest observations.
\item Verifying the broad outlines of the White--Rees~(1978) picture for
baryon condensation within dark haloes.

\end{itemize}

\section{The Cutting Edge: A Case Study}

The challenge of simulating the formation of galaxies in the
cosmological context requires huge computational resources. The
estimates given above suggest 10$^8\hbox{--}9$ particles. Both the
computer time and the amount of memory required for such large
simulations pose severe computational constraints and, at present, are
available only on massively parallel supercomputers. A number of
groups are working in this area with parallel codes. I will describe
the ``Virgo'' collaboration in which I am involved (Pearce et
al.~1996). This is a group of eight, primarily UK researchers, using
the UK national supercomputer (a Cray T3D) in Edinburgh as well as a
Cray T3E in Munich.

We have ported the AP$^3$M code to the Edinburgh 512 processor Cray
T3D using CRAFT, Cray's proprietary software. This software creates a
global address space which appears flat to the programmer vastly
easing the difficulty of parallelizing code for a distributed memory
architecture. The penalty, however, is a certain amount of
inefficiency and lack of portability. For this reason an explicit
message passing (MPI) version is now being developed.

The primary difficulty with parallel N-body codes is to achieve an
efficient data and work distribution. With grid codes there is the
added difficulty that the mapping from a perhaps highly clustered
particle distribution to a regularly spaced grid may lead to a large
communications overhead. Tree codes tend to be somewhat better behaved
in this regard as they avoid the potential many-to-one mapping of the
Lagrangian particles to the Eulerian grid and because of the use of a
more flexible data structure. In general, however, care must be taken
with particle codes to minimize the communication overhead. The
particle is the fundamental element and essentially no work is done
without reference to its neighbours, therefore data must be carefully
distributed to avoid costly inter-processor communication. This
behaviour can be compared with that in finite element codes in which a
large amount of work is done internally to the element and the ratio
of computation to communication is much higher.

The parallel AP$^3$M refined grid code scales well up to 256
processors, although some load imbalance occurs for heavy
clustering. We have used a strategy in which we distribute data in one
spatial direction over the processors. Provided there are a number of
particle clusters present in the simulation volume this is sufficient
to average out the worst load imbalances. We are now achieving
performances of 2500 particles/s/T3D node for collisionless
simulations and 700 particles/s/T3D node for the AP$^3$M-SPH code
``Hydra'' (Couchman et al.~1995). A $2\times10^7$ particle run with
5000 steps would thus take approximately one week on 256 nodes of a Cray
T3D; corresponding to 5 years total processing time. This number of
particles uses about half the 64Mb available per node on 256 nodes. A
Cray T3E has 3 times faster processors and can have up to 2Gb of
memory per node; 10$^8$ particles should be feasible.

\section{The Future}

I will conclude by making a few remarks about the future of
cosmological simulations. It is clear that the trend towards larger
simulations will continue. As the dynamic range of a simulation
increases so does the range of density contrasts that can in principle
be modelled; alternatively, the first structures which we can
confidently identify in a simulation form at earlier epochs. Accurate
modelling demands that the simulation can deal with the wide range of
densities and timescales which arise. Both spatial and temporal
adaptivity will become more important. Tree codes frequently have both
whilst the AP$^3$M code lags, for example, in having only spatial
adaptivity.

One of the most obvious problems with large
supercomputer applications is the vast amount of data that is
generated. Cosmological N-body codes are no exception. Each timeslice
of a $2\times10^7$ particle run occupies 960Mb (64bit) for position
and velocity storage. The output of a hydrodynamic code would be
nearly 50\% greater---$\sim1.5$\thinspace Gb per timeslice.  

An approach which helps avoid the data explosion is to concentrate on
selected volumes. This has generated a great deal of interest recently
with a large number of workers pursuing different avenues of
attack. There is now a general appreciation that external fields must be
carefully applied to the selected volumes to properly mimic the tidal
effects of the external universe. This general approach is very
promising. It remains to be seen whether the information gained from
these studies will be sufficient to allow us to predict the locations
of galaxies in larger-volume (perhaps collisionless) simulations. If
we are to properly assess the upcoming large sky surveys, and hence to
be able to understand galaxy distributions, morphologies, bias etc.,
we must be able to generate reliable theoretical galaxy catalogues
from simulations.

\end{document}